\documentclass[journal]{IEEEtran}
\ifCLASSINFOpdf
\else
\fi
\usepackage{amsfonts}
\usepackage{amsmath,amsfonts,amssymb,epsfig,subfigure,cite,url,multirow,booktabs}
\usepackage{afterpage,mathrsfs,algorithm,algorithmic}
\usepackage{amsthm}

\newtheorem{lemma}{\textbf{Lemma}}

\newtheorem{proposition}{\textbf{Proposition}}
\newtheorem*{remark}{\textbf{Remark}}

\begin{document}
%
\title{RIS-Assisted Wireless Link Signatures for Specific
Emitter Identification: Preliminary}
%
%
%


\author{Ning Gao,~\IEEEmembership{Member,~IEEE,}~Shuchen Meng,~Cen Li,~Shengguo Meng,~Wankai Tang,~\IEEEmembership{Member,~IEEE,}\\~Shi Jin,~\IEEEmembership{Fellow,~IEEE,} and~Michail Matthaiou,~\IEEEmembership{Fellow,~IEEE}

\thanks{N. Gao, S. Meng and C. Li are with the School of Cyber Science
and Engineering, Southeast University, Nanjing 210096, China (e-mail:
ninggao@seu.edu.cn).}
\thanks{S. Meng, W. Tang and S. Jin are with the National Mobile Communications
Research Laboratory, Southeast University, Nanjing 210096, China (e-mail: seumengsg@seu.edu.cn; tangwk@seu.edu.cn; jinshi@seu.edu.cn).}
\thanks{
M. Matthaiou is with the Centre for Wireless Innovation (CWI), Queen’s University Belfast, Belfast BT3 9DT, U.K. (e-mail: m.matthaiou@qub.ac.uk).}
}
%
%

\markboth{UNSUBMITTED DRAFT}%
{Shell \MakeLowercase{\textit{et al.}}: Bare Demo of IEEEtran.cls for IEEE Journals}
%



\maketitle

\begin{abstract}
The specific emitter identification (SEI) is a promising technology which can enhance the access security of a massive number of devices in the near future. In this paper, we propose a reconfigurable intelligent surface (RIS)-assisted SEI system, in which the legitimate transmitter can customize the channel fingerprints during SEI by controlling the ON-OFF state of the RIS. Without loss of generality, we use the received
signal strength (RSS) based spoofing detection approach to analyze the feasibility of the proposed architecture. Specifically, based on the RSS, we derive the statistical properties of SEI and give some interesting insights, which showcase that the RIS-assisted SEI is theoretically feasible. Then, we derive the optimal detection threshold to maximize the performance in the context of the presented performance metrics. Next, the actual feasibility of the proposed system is verified via proof-of-concept experiments on a RIS-assisted SEI prototype platform. The experiment results show that there are 3.5\% and 76\%  performance improvements when the transmission sources are at different locations and at the same location, respectively.
\end{abstract}


%
\IEEEpeerreviewmaketitle
\section{Introduction}
\IEEEPARstart{T}{hanks} to the rapid development of the sixth-generation (6G) wireless
networks, we are gradually moving form the Internet of Things (IoT) era into the Internet of everything era. However, future heterogeneous IoT networks will face the formidable challenge of securing the access of an increasing number of devices. Recently, it has been proposed that zero-trust security should be continuously verified and validated at every access stage of a IoT device before granting admission to network resources \cite{10032070}. Therefore, whether it is the traditional high-level authentication or the contemporary zero-trust security, the overhead of the trusted access will be anyway enormous for a massive number of dynamic connections. The specific emitter identification (SEI) is a promising technology to authenticate different transmitters from the wireless signal propagation perspective. Generally, the fingerprints used for SEI include the channel fingerprints and radio frequency (RF)
fingerprints, such as the channel frequency response
(CFR), received signal strength (RSS) or the carrier frequency offset (CFO) and input/output (I/Q) imbalance, etc \cite{7539590}. Since the fingerprints come from the endogenous products of the communication process, the SEI has the advantages of low latency, low overhead and good compatibility. As a result, the SEI can indeed play a pivotal role in the 6G security exercise.

Compared to the RF fingerprints, the channel fingerprints are more convenient to acquire because they do not require sophisticated hardware and strict time synchronization \cite{7539590}. In one of the early works in this space, the spatial variability of CFR has been utilized to authenticate the transmitter in a typical indoor environments \cite{4533330}. Considering the spoofing and sybil attacks, the authors of \cite{5427107} proposed a RSS based authentication technology to detect and locate the attacks. To enhance the performance, the channel fingerprints power enabled multi-user collaborative SEI was investigated \cite{7839273}. Apart from the above mentioned, the multi-observation/multi-attribute have been considered to provide a robust authentication performance by introducing multi-dimension fingerprints information \cite{8234675,8533399,9340570}. With the development of artificial intelligence (AI), AI-assisted SEI mechanisms have been gradually emerging, by leveraging Bayesian classifiers, extreme learning machines, deep learning and reinforcement learning \cite{8961122,GAO,7891506}. In virtue of the fine-grained feature extraction and the strong nonlinear learning abilities of AI, these methods can achieve a satisfactory performance in specific wireless environments. However, the resilience of SEI for dynamic wireless environment is extremely fragile. The channel fingerprints and noise reflect the inherent properties of the wireless propagation environments, which cannot be controlled artificially. Regardless of the traditional SEI or AI-assisted SEI, when such inherent properties are not conducive to authenticate different transmitters, the performance of the SEI is difficult to guarantee. Furthermore, one big problem of channel fingerprints based SEI is that the emitters to be identified should be
potentially located at spatially separated positions \cite{4533330}. So far, this problem has not been well addressed in the related literature.
\subsection{Motivation and Contributions}
A reconfigurable intelligent surface (RIS) is a planar digitally programmable metasurface, which can manually customize wireless propagation environments via a programmable field-programmable gate array (FPGA) controller \cite{9424177}. For this reason, the evolution of RISs
has spawned a number of related academic contributions in wireless communication and security over the past few years. To name but a few, a RIS can act as a signal reflection hub to support passive beamforming, coverage enhancement and wireless sensing, etc \cite{8746155,9462487,9326394}. Form a secrecy rate perspective, reference \cite{9206080} considered the RIS-aided wireless secure communication, where a passive RIS is utilized to guarantee the system secrecy rate. For the key-based physical layer security (PLS), a RIS can be used for assisting physical layer key generation in low-entropy environments \cite{ARXIV22GAO}. For instance, secret key generation schemes with RISs have been proposed, where the dynamic channel environment is constructed \cite{9442833} and the RIS reflecting coefficients are optimized to maximize the sum secret key rate \cite{9663196}. Very recently, RIS-assisted PLS is in full swing, however, the great potential of RIS for SEI has not yet attracted an equal amount of research attention. To increase the resilience of SEI, in this paper, for the first time, we propose a new paradigm of RISs, namely, the RIS-assisted SEI. The feasibility of RIS-assisted SEI system is analyzed from a theoretical perspective and real-world experiments. Our main contributions are summarized as follows:
\begin{itemize}
  \item To the best of the authors' knowledge, the proposed RIS-assisted SEI system is one of the first works that employ the RIS for SEI. The passive nature of RISs, that have no radio frequency chains and signal amplification function, helps avoiding introducing
additional thermal noise into the channel fingerprints. More importantly, the proposed new architecture gives a potential solution to the long-standing problem that the transmitters must be located at different spatial locations for the channel fingerprints based SEI.
  \item Without loss of generality, we adopt the RSS based spoofing detection approach to analyze the feasibility of the proposed RIS-assisted SEI system. In particular, we derive the statistical properties of RIS-assisted SEI via RSS, including the asymptotic distributions of RSS and RSS statistical distance, respectively. Then, some interesting insights are given based on the statistical properties. Moveover, we give the metrics to evaluate the performance of the proposed system and derive the optimal detection threshold to maximize the performance.
  \item To verify the proposed RIS-assisted SEI system, we develop a prototype platform and conduct proof-of-concept experiments. The experiment results show that there are 3.5\% and 76\%  performance improvements when the transmission sources are at different locations and at the same location, respectively. Last but not least, the proposed RIS-assisted SEI system is compatible with the existing RIS-assisted communications in protocols and hardware. Thus, this can spur new research areas to design integrating communications and security (ICAS) by sharing RISs for trusted access and communications \cite{ARXIV22GAO}.
\end{itemize}
\subsection{Notations}
The scalars are in italic letters, whilst vectors and matrices are given by bold-face lower-case and
upper-case letters, respectively. The operator $|\cdot|$ denotes Euclidean norm and $\text{diag}[\cdot]$ denotes a diagonal matrix. The operators $(\cdot)^*$ and $(\cdot)^H$ represent the conjugate and conjugate transpose, respectively. The operators $\mathbb{E}[\cdot]$ and  $\mathbb{D}[\cdot]$ denote the statistical expectation and variance, respectively. Finally, $\text{Re}\{{\cdot}\}\sim \mathcal{N}(\cdot,\cdot)$ is the real part of a complex number following a Gaussian distribution.

\section{system model and problem formulation}\label{sec:2}

\subsection{Channel Model}
We consider a static three components RIS-assisted SEI system that consists of the RIS, the legitimate users Alice and Bob, and the malicious user Eve, which are denoted as $R$, $A$, $B$ and $E$, respectively.
All of the participants are resource-limited devices, which are equipped with a single antenna. Alice intends to access Bob and transmits the wireless data via orthogonal frequency division
multiplexing (OFDM), while Eve tries to mimic the identity of Alice to access Bob and transmits the illegal data. Due to the regulations of the access control protocol, users cannot send access requests simultaneously. In particular, Alice attempts to improve the SEI performance by controlling the ON and OFF states of RIS. When Alice requests to access Bob, the RIS is in the ON state to reflect the wireless signal. Then, the received signal at Bob on the $k$th subcarrier of the $n$th OFDM symbol can be denoted as
\begin{equation}
\label{eq:A2B}
y_{B,k}[n]=\big(\overbrace{\underbrace{\mathbf{h}^T_{RB,k}\mathbf{\Phi}_k\mathbf{h}_{AR,k}}_{\text{Cascaded channel}}
+h_{AB,k}}^{\text{Equivalent
channel}}\big)x_{A,k}[n]+z_k[n],
\end{equation}
where $k\in\{1,\ldots,K\}$ is the subcarrier number of the OFDM symbol, $\mathbf{h}_{AR,k}\in\mathbb{C}^{N\times1}$ is the channel gain from Alice to RIS, $\mathbf{\Phi}_k=\text{diag}\big[\alpha_1e^{j\theta_{1,k}},\alpha_2e^{j\theta_{2,k}},\ldots,\alpha_Ne^{j\theta_{N,k}}\big]$ is the RIS diagonal phase-shift matrix with $N$ reflection elements and the amplitude coefficient $0\leq\alpha_n\leq1, n\in\{1,\ldots,N\}$, $\mathbf{h}_{RB}\in\mathbb{C}^{N\times1}$ is the channel gain from the RIS to Bob, $h_{AB,k}\in\mathbb{C}^{1\times1}$ is the channel gain of the direct link from Alice to Bob, $x_{A,k}[n]$ is the data symbol of the $k$th subcarrier and $z_k[n]$ is the noise at Bob following a zero-mean complex Gaussian distribution with variance $\sigma^2$. When Alice has no access request, the RIS is in the OFF state and does not reflect wireless signals. Thus, if Eve tries to access Bob, the received signal at Bob on the $k$th subcarrier of the $n$th OFDM symbol is given by
\begin{equation}
\label{eq:B2A}
y_{B,k}[n]=h_{EB,k}x_{E,k}[n]+z_k[n],
\end{equation}
where $h_{EB,k}\in\mathbb{C}^{1\times1}$ is the channel gain of the direct link from Eve to Bob, and $x_{E,k}[n]$ is the illegal data symbol of the $k$th subcarrier. We consider the narrowband channel fading model, where $\mathbf{h}_{AR,k}$ and $\mathbf{h}_{RB,k}$ follow an independent identically distributed (i.i.d.) zero-mean complex Gaussian distribution with covariance matrices $\sigma_{AR}^2\mathbf{I}_N$ and $\sigma_{RB}^2\mathbf{I}_N$, respectively. The channel $h_{AB}$ follows a zero-mean complex Gaussian distribution with variance $\sigma_A^2$. Similarly, the channel $h_{EB}$ follows a zero-mean complex Gaussian distribution with variance $\sigma_E^2$.
\subsection{Spoofing Detection Formulation}
Suppose that we are interested in the received $n$th OFDM symbol across the $k$th subcarriers at the sampling frequency $f_s$. For simplicity of exposition, we remove the symbol $[n]$ and the subscript $k$ in the following. Let $\tau$ be the available detection time, and then the discrete sampled signal at Bob is represented as $\mathbf{y}_B=[y_B(1),\ldots,y_B(L)]$ with the number of samples $L=\tau f_s$. Thus, we formulate the SEI as a spoofing detection problem, and the RSS can be calculated by
\begin{equation}
T(y_B)=\frac{1}{L}\mathbf{y}_B\mathbf{y}^H_B.
\end{equation}
When $N$ is fixed and $L$ goes to infinity, $T(y_B)$ can approximate the statistical property of the RSS. Then, the statistical significance testing of the spoofing detection is formulated as,
\begin{align}
&\mathcal{H}_0:\text{normal (no attack)},\nonumber\\
&\mathcal{H}_1:\text{abnormal (under attack)},\nonumber
\end{align}
and the test statistic for spoofing detection is represented as the RSS statistical distance between the observation and the legal reference, which is denoted as
\begin{equation}
\label{eq:tests}
\Delta T(y_B)=T(y^{\text{ref}}_B)-T(y^{\text{obs}}_B).
\end{equation}
In significance testing, the test statistic $\Delta T(y_B)$ is utilized to evaluate the transmission source of the observation $y_B^{\text{obs}}(l),l\in\{1,\ldots,L\}$. For a detection threshold $\epsilon$, we define an acceptance region $\Omega$ and a critical region $\Omega^c$. Then, we declare the $\mathcal{H}_0$ hypothesis valid (no attack) if the test statistic $\Delta T(y_B)\in\Omega$ and the $\mathcal{H}_1$ hypothesis valid (under attack) if the test statistic $\Delta T(y_B)\in\Omega^c$.

\section{Theoretical Analysis of Spoofing Detection}\label{sec:3}

\subsection{Test Statistics for Spoofing Detection}
Under hypothesis $\mathcal{H}_0$, the received signal is rewritten as
\begin{equation}
\label{eq:rA2B}
y_{B}=\bigg(\sum_{n=1}^N\alpha_ne^{j\theta_n}h_{RB,n}h_{AR,n}
+h_{AB}\bigg)x_{A}+z.
\end{equation}
\begin{lemma}
Suppose that the equivalent channel of the $k$th subcarrier from Alice to Bob is
\begin{equation}
h_{ARB}=\sum_{n=1}^N\alpha_ne^{j\theta_n}h_{RB,n}h_{AR,n}
+h_{AB,n},
\end{equation}
then, the mean and the variance of the equivalent channel $h_{ARB}$ is zero and $\sum_{n=1}^N\alpha^2_ne^{j2\theta_n}\sigma^2_{RB}\sigma^2_{AR}+\sigma^2_A$, respectively.
\end{lemma}
\begin{IEEEproof}
Since $\forall n\in\{1,\ldots,N\}$, the distribution of the channels $h_{RB,n}$ and $h_{AR,n}$ follow an i.i.d., hence, the statistical expectation of $h_{ARB}$ can be given by
\begin{align}
\mu_h=\mathbb{E}[h_{ARB}]&=\sum_{n=1}^N\alpha_ne^{j\theta_n}\mathbb{E}[h_{RB}]\mathbb{E}[h_{AR}]
+\mathbb{E}[h_{AB}]=0.\nonumber
\end{align}
Moreover, the variance of $h_{ARB}$ can be calculated by
\begin{align}
\sigma_h^2
&=\sum_{n=1}^N\alpha^2_ne^{j2\theta_n}\mathbb{E}[|h_{RB}|^2]\mathbb{E}[|h_{AR}|^2]+\mathbb{E}[|h_{AB}|^2]\nonumber\\
&=\sum_{n=1}^N\alpha^2_ne^{j2\theta_n}\sigma^2_{RB}\sigma^2_{AR}+\sigma^2_A.\nonumber
\end{align}
Then, the proof is completed.
\end{IEEEproof}
Since the components of $h_{ARB}$ include the product of $h_{RB}$ and $h_{AR}$, it is quite challenging to obtain the probability density functions (PDF) of $h_{ARB}$ \cite{9079918}. The random variables $y_B(l), l\in\{1,\ldots,L\}$, $h_{RB}$ and $h_{AR}$, are i.i.d., whilst $T(y_B)$ and $h_{ARB}$  are the sum of i.i.d. random variables, which motivates us to characterize the statistical properties of the RSS by using the central limit theorem (CLT).
\begin{proposition}
Under hypothesis $\mathcal{H}_0$, for a large $L,N$, the PDF of $T(y_B)$ can be approximated by a Gaussian distribution with mean $\mu_0=\sum_{n=1}^N\alpha^2_ne^{j2\theta_n}\sigma^2_{RB}\sigma^2_{AR}+\sigma^2_A+\sigma^2$,
and variance
\begin{align}
\label{eq:var0}
\sigma^2_0=\frac{1}{L}\bigg(2\sigma_h^4+2\sigma^4-\mu^2_0\bigg).
\end{align}
\end{proposition}
\begin{IEEEproof}
The proof is given in Appendix \ref{sec:A}.
\end{IEEEproof}
Following a similar procedure, for a large $L$, we can derive that the PDF of $T(y_B)$ under hypothesis $\mathcal{H}_1$ follows a Gaussian distribution with mean $\mu_1=\sigma^2_E+\sigma^2$ and variance $\sigma^2_1=\frac{1}{L}\bigg(2\sigma_E^4+2\sigma^4-\mu^2_1\bigg)$. Based on the PDFs of $T(y_B)$ under hypothesis $\mathcal{H}_0$ and $\mathcal{H}_1$, respectively, the PDF of $\Delta T(y_B)$ in \eqref{eq:tests} can be given in the following proposition.
\begin{proposition}
Under hypothesis $\mathcal{H}_0$, the PDF of the test statistic $\Delta T(y_B)$ follows a Gaussian distribution with mean $\mu_{\Delta0}=0$ and variance
\begin{align}
\label{eq:deltav0}
\sigma^2_{\Delta0}=\frac{2}{L}\bigg(2\sigma_h^4+2\sigma^4-\mu^2_0\bigg).
\end{align}
On the other hand, under hypothesis $\mathcal{H}_1$, the PDF of the test statistic $\Delta T(y_B)$ follows a Gaussian distribution with mean $\mu_{\Delta1}=\sum_{n=1}^N\alpha^2_ne^{j2\theta_n}\sigma^2_{RB}\sigma^2_{AR}+\sigma^2_A-\sigma_E^2$ and variance
\begin{align}
\label{eq:deltav1}
\sigma^2_{\Delta1}=\frac{1}{L}\bigg(2\sigma_h^4+2\sigma_E^4+4\sigma^4
-\mu^2_0-\mu^2_1\bigg).
\end{align}
\end{proposition}
\begin{IEEEproof}
From \eqref{eq:tests}, the statistical expectation of $\Delta T(y_B)$ can be written as
\begin{align}
\label{eq:meandelta}
\mathbb{E}[\Delta T(y_B)]=\mathbb{E}[T(y^{\text{obs}}_B)]-\mathbb{E}[T(y^{\text{ref}}_B)],
\end{align}
and the variance of $\Delta T(y_B)$ can be written as
\begin{align}
\label{eq:vardelta}
\mathbb{E}[(\Delta T(y_B)-\mathbb{E}[\Delta T(y_B)])^2]~~~~~~~~~~~~~~~~~~~~~~~~~~~~~~~~~~~~\nonumber\\
=\mathbb{E}[(T(y^{\text{obs}}_B))^2]-\mathbb{E}[T(y^{\text{obs}}_B)]^2+\mathbb{E}[(T(y^{\text{ref}}_B))^2]-\mathbb{E}[T(y^{\text{ref}}_B)]^2\nonumber\\
=\mathbb{D}[T(y^{\text{obs}}_B)]+\mathbb{D}[T(y^{\text{ref}}_B)].~~~~~~~~~~~~~~~~~~~~~~~~~~~~~~~~~~~~~~~~
\end{align}
By substituting $\mu_0$, $\sigma^2_0$, $\mu_1$ and $\sigma^2_1$  into \eqref{eq:meandelta} and \eqref{eq:vardelta}, we can complete the proof.
\end{IEEEproof}

\begin{remark}
Regarding the PDF of $\Delta T(y_B)$ for RIS-assisted SEI, we obtain the following interesting insights:
1) First and foremost, the long-standing assumption that the transceiver should be located at different spatial locations can be relaxed. Specifically, from the PDFs of $\Delta T(y_B)$ under hypothesis $\mathcal{H}_1$, it can be seen that the mean $\mu_{\Delta1}$ is bigger than that without RIS. Intuitively, it shows that the PDF of $\Delta T(y_B)$ under hypothesis $\mathcal{H}_1$ moves horizontally to the right side, compared with no RIS assistance. This phenomenon implies that the RSS statistical distance between the hypothesis $\mathcal{H}_0$ and the hypothesis $\mathcal{H}_1$ is enlarged by the assistance of RIS. 2) Obviously, we can find that the amplitude coefficient $\alpha_n$ and the phase-shift $\theta_n$ can be configured to maximize the RSS statistical distance. In other words, the cascaded channel of the RIS can introduce a maximum equivalent channel gain $N\sigma^2_{RB}\sigma^2_{AR}+\sigma^2_A$ with the amplitude coefficient $\alpha_n=1$ and the coherent phase-shift. Furthermore, we see that the size of the RIS can be optimized to approach an optimal performance. However, the RIS may also increase the variance of the equivalent channel which is destructive to SEI, thus, there is a fundamental trade-off behind the optimization of the RIS structure.
\end{remark}
\subsection{Metrics}
We present the metrics for evaluating the performance of the proposed RIS-assisted SEI system, which includes the probability of detection, and the probability of false alarm. The probability of detection can be defined as the percentage of attack attempts that
are determined to be under attack. Under hypothesis $\mathcal{H}_1$, the probability of detection is expressed as
\begin{align}
\label{eq:pd}
P_d=\int_{\epsilon}^{+\infty}p_{\Delta1}(s)ds=Pr(\Delta T(y_B)>\epsilon|\mathcal{H}_1).
\end{align}
Based on Proposition 2, the probability of detection can be further written as
\begin{align}
P_d=Q\bigg(\frac{\epsilon-\mu_{\Delta1}}{\sigma_{\Delta1}}\bigg),
\end{align}
where $\mathcal{Q}(\cdot)$ is the complementary distribution function of the standard Gaussian function, i.e., $\mathcal{Q}(s)=\frac{1}{\sqrt{2\pi}}\int_{s}^{+\infty}e^{-\frac{t^2}{2}}dt$.
Similarly, the probability of false alarm corresponds to the probability of declaring a false positive under hypothesis $\mathcal{H}_0$, which is defined as
\begin{align}
\label{eq:pf}
P_f=\int_{\epsilon}^{+\infty}p_{\Delta0}(s)ds=Pr(\Delta T(y_B)>\epsilon|\mathcal{H}_0).
\end{align}

For a target probability of false alarm, i.e., $0.1$, the detection threshold can be determined by $\epsilon=\mathcal{Q}^{-1}(P_f)\sigma_{\Delta0}+\mu_{\Delta0}$. Furthermore, to detect a spoofing attack as accurately as possible, we derive the optimal detection threshold in Proposition 3.
\begin{proposition}
The optimal detection threshold of the RIS-assisted SEI system by using RSS, that minimizes the probability of erroneous detection, is given by
\begin{align}
\label{eq:threshold}
\epsilon^\star=~~~~~~~~~~~~~~~~~~~~~~~~~~~~~~~~~~~~~~~~~~~~~~~~~~~~~~~~~~~~~~~~\nonumber\\
\frac{\sigma^2_{\Delta0}\mu_{\Delta1}+\sqrt{\sigma^2_{\Delta0}\sigma^2_{\Delta1}\mu^2_{\Delta1}+2\sigma^2_{\Delta0}\sigma^2_{\Delta1}(\sigma^2_{\Delta0}-\sigma^2_{\Delta1})\ln\frac{\sigma_{\Delta0}}{\sigma_{\Delta1}}}}{\sigma^2_{\Delta0}-\sigma^2_{\Delta1}}.
\end{align}
\end{proposition}
\begin{IEEEproof}
See Appendix \ref{sec:B}.
\end{IEEEproof}
\section{Experimental Results}\label{sec:4}
\subsection{Experimental Setup}
A RIS-assisted SEI prototype platform is developed to collect RSS data from different transmitters. As shown in Fig. \ref{Fig:PLA1}, the developed prototype platform consists of the high-performance notebook HOST PCs for signal processing, the software radio platform USRP-RIO with the synchronous clock node WR LEN and the clock distributor WR switch to transmit and receive wireless signals, and a RIS for assisting the SEI. We conduct two experiments to verify the feasibility and the performance:
\begin{figure}[!ht]
 \centering
  \includegraphics[width=8cm]{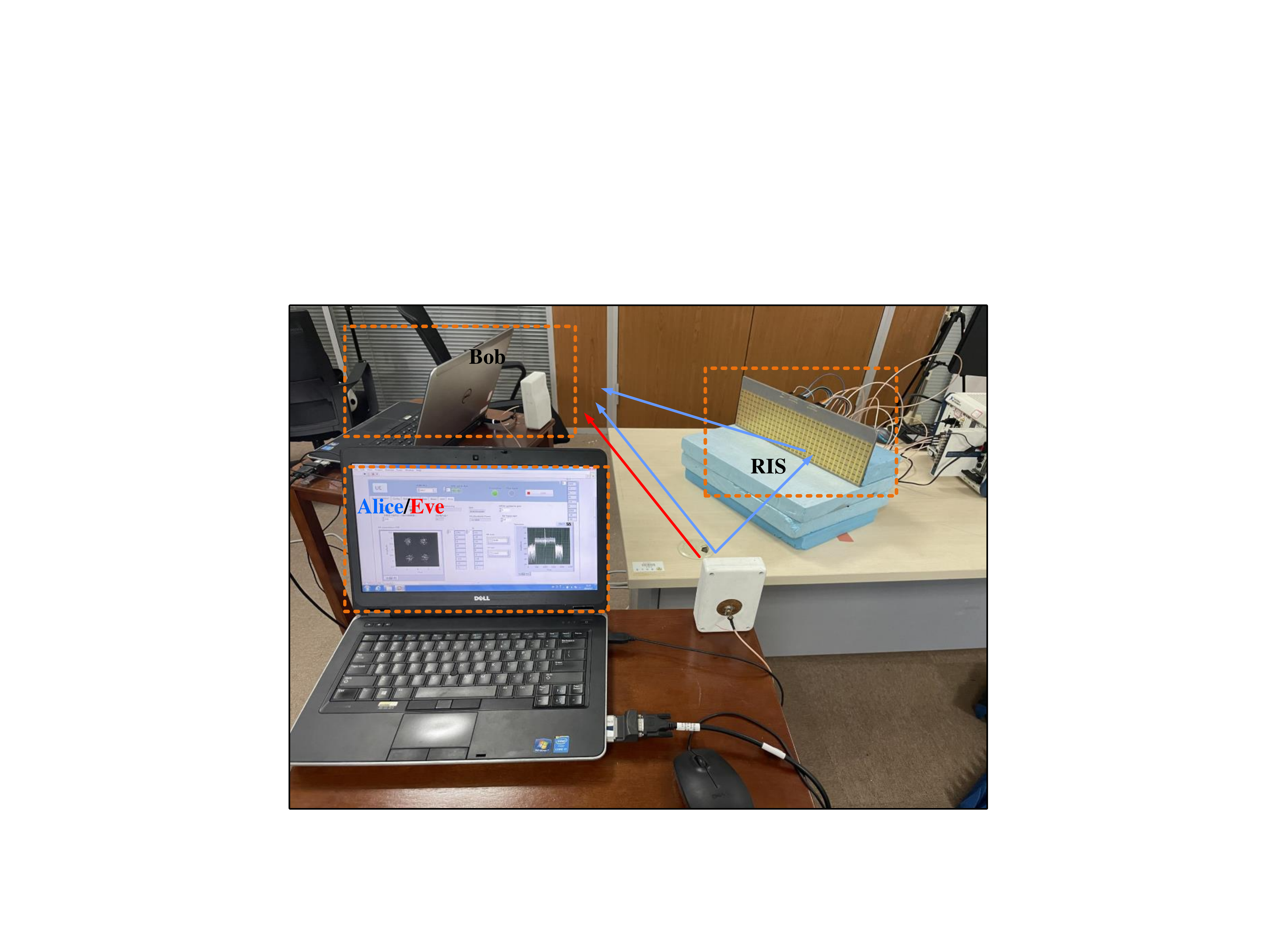}
  \caption{The schematic diagram of the developed RIS-assisted SEI platform.}
  \label{Fig:PLA1}
\end{figure}
\begin{itemize}
  \item \textbf{Experiment 1}: Alice and Eve are placed at different locations, where Bob is 2 m away from Alice and Eve is 4 m away from Alice. We use ``with RIS" and ``without RIS" to represent the states ``ON" and ``OFF", respectively. The goal of this experiment is to verify the constructive role of RIS in SEI.
  \item \textbf{Experiment 2}: Alice and Eve at the same position, where  Eve takes Alice's place when Alice leaves, and the distance is 2 m away from Bob. The purpose of this experiment is to validate the effectiveness of RIS on solving the long-standing SEI problem that the transmitters must be located at different spatial locations.
\end{itemize}
\subsection{Performance Analysis}

\begin{figure}[!ht]
 \centering
  \includegraphics[width=8cm]{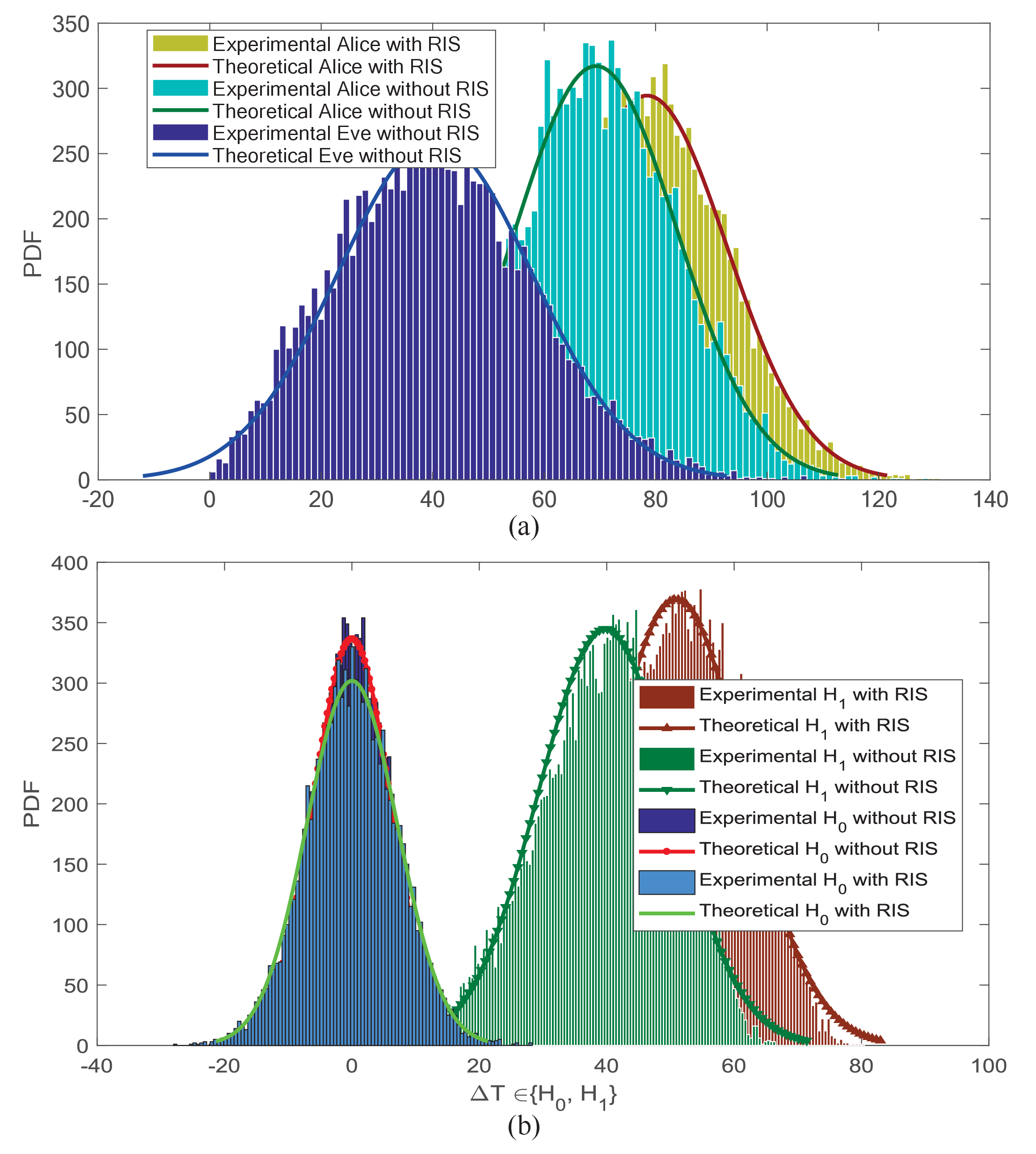}
  \caption{The various statistical distributions of different transmission sources.}
  \label{Fig:PLADIS}
\end{figure}

Figure \ref{Fig:PLADIS} illustrates the various statistical distributions of different transmitters when the distance is set based on Experiment 1. We can find that the statistics of RSS and the RSS distance can be well approximated by Gaussian distributions. Figure \ref{Fig:PLADIS} (a) shows that the overlap of the RSS distribution decreases with the assistance of RIS.
This means that the channel fingerprints of Alice and Eve have greater individual feature than that without RIS. The deployment of RIS adds natural endogenous properties to the wireless propagation environments. Here, the added endogenous properties are reflected in the mean and variance of the RSS distribution. In other words, for a given thermal noise, we observe that the increased signal-to-noise-rate (SNR) of the legitimate user can improve the SEI performance. From Fig. \ref{Fig:PLADIS} (b), we observe that the PDF of $\Delta T$ in hypothesis $\mathcal{H}_0$ follows a zero mean Gaussian distribution and the variance with RIS is higher that without RIS. Obviously, when Alice switches on RIS during the SEI, we see that the PDF overlap of $\Delta T$ between hypothesis $\mathcal{H}_0$ and hypothesis $\mathcal{H}_1$ becomes smaller. The overall error probability of declaring that a test statistic belongs to hypothesis $\mathcal{H}_0$ and hypothesis $\mathcal{H}_1$ is reduced. That is, the probability of detection increases and the probability of false alarm decreases with the assistance of RIS. These results also corroborate the validity of the theoretical analysis.
\begin{figure}[!ht]
 \centering
  \includegraphics[width=8cm]{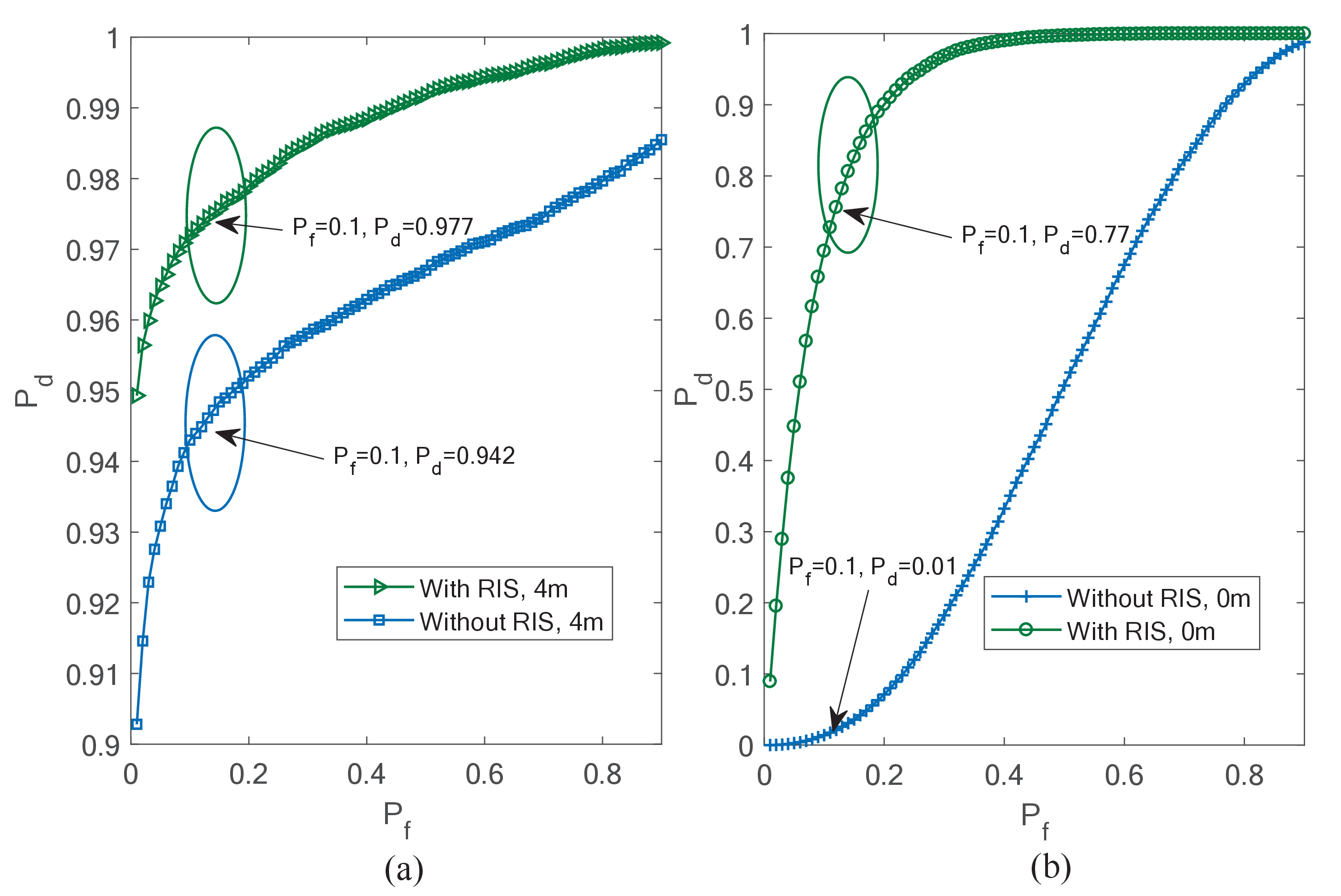}
  \caption{The ROC curves of the proposed RIS-assisted SEI system and the traditional SEI system with different locations of Eve.}
  \label{Fig:ROC}
\end{figure}

As shown in Fig. \ref{Fig:ROC}, we plot the receiver operating characteristic (ROC) curves to evaluate the proposed RIS-assisted SEI system. From Fig. \ref{Fig:ROC} (a), we can find that when the distance between Eve and Bob is 4 m and the target probability of false alarm is $P_f=0.1$,
the probability of detection is $P_d=0.942$ without RIS, whilst the probability of detection is $P_d=0.977$ with RIS. This observation indicates that the probability of detection yields a 3.5\% improvement. It is noted that there is a significant improvement when the scale of access devices is large. The worse the channel quality of Eve, the more significant the SEI performance gain brought by the RIS. It is interesting to check the extreme scenario, where the channel quality of Eve is as good as Alice's, implying that Alice and Eve are at the same location. In this case, we utilize the traditional channel fingerprints based SEI as a benchmark and analyze the performance of the proposed RIS-assisted SEI system, which is shown in Fig. \ref{Fig:ROC} (b). The traditional channel fingerprints based SEI cannot detect the spoofing, which is located at the same position of the legitimate user; thus, the probability of detection is only $P_d=0.01$ with $P_f=0.1$. Interestingly, with the assistance of the RIS, the probability of detection approaches $P_d=0.77$. Compared to without RIS, there is a significant increase of 76\% of the detection probability with RIS.

\section{Conclusion}\label{sec:5}
In this paper, we have proposed a new RIS-assisted SEI system, in which the legitimate transmitter can customize the channel fingerprints during SEI by controlling the ON-OFF state of the RIS. We have theoretically analyzed the feasibility of the proposed system followed by proof-of-concept experiments. On this basis, we have given some interesting insights to guide the future design of the RIS-assisted SEI. The experiments show that there is 3.5\% performance improvement when the transmission sources are at different locations and a significant 76\% performance improvement when the transmission sources are at the same location. Most importantly, the long-standing problem for channel fingerprints based SEI can be solved by using the proposed RIS-assisted SEI system.
\appendices
\section{Proof of Proposition 1}\label{sec:A}
By invoking the CLT, when $L$ is large enough, the mean of $T(y_B)$ is equal to the sample mean, which is given by
\begin{align}
\mu_0=\mathbb{E}[T(y_B)]=\sum_{n=1}^N\alpha^2_ne^{j2\theta_n}\sigma^2_{RB}\sigma^2_{AR}+\sigma^2_A+\sigma^2.
\end{align}
Then, the variance of $T(y_B)$ is calculated as follows
\begin{align}
\label{eq:varDe0}
\sigma^2_0
&=\mathbb{E}\bigg[\bigg(\frac{1}{L}\sum_{l=1}^L|y_B(l)|^2-\mu_0\bigg)^2\bigg]\nonumber\\
&=\frac{1}{L}\mathbb{E}\bigg[\bigg(|h_{ARB}(l)|^2+|z(l)|^2+h_{ARB}(l)z^*(l)\nonumber\\
&+h_{ARB}^*(l)z(l)-\mu_0\bigg)^2\bigg].
\end{align}
On the other hand, when $N$ is very large, it can be shown that the cascaded
channel of the RIS can be approximated as
a complex Gaussian random variable \cite{9079918}. Given that $h_{ARB}(l)$ and $z(l)$ are i.i.d. and circularly symmetric, i.e., $\text{Re}\{h_{ARB}\}\sim \mathcal{N}\bigg(0,\frac{\sum_{n=1}^N\alpha^2_ne^{j2\theta_n}\sigma^2_{RB}\sigma^2_{AR}+\sigma^2_A}{2} \bigg)$, we can get $\mathbb{E}[(h_{ARB}(l))^2]=0$ and $\mathbb{E}[(z(l))^2]=0$ \cite{4373414}. By substituting  $\mathbb{E}[|h_{ARB}(l)|^4=2\sigma_h^4$ and $\mathbb{E}[|z(l)|^4=2\sigma^4$ into \eqref{eq:varDe0}, the variance of $T(y_B)$ can be derived as \eqref{eq:var0}, and then, the proof is completed.
\section{Proof of Proposition 3}\label{sec:B}
The probability of erroneously detecting the $\Delta T(y_B)$ under $\mathcal{H}_0$ as $\Delta T(y_B)$ under $\mathcal{H}_1$, is equal to the probability of false alarm $P_f$, and we define it as $\varepsilon_{\Delta0}$. Similarly, the probability of erroneously detecting $\Delta T(y_B)$ under $\mathcal{H}_1$ as $\Delta T(y_B)$ under $\mathcal{H}_0$, is equal to the probability of miss detection $P_m=1-P_d$, and we define it as $\varepsilon_{\Delta1}$.
With \eqref{eq:pd}, \eqref{eq:pf}, the overall probability of detection error is given by
\begin{align}
\varepsilon(\epsilon)&=\varepsilon_{\Delta0}+\varepsilon_{\Delta1}\nonumber\\
&=1-\int_{\epsilon}^{+\infty}p_{\Delta1}(s)ds+\int_{\epsilon}^{+\infty}p_{\Delta0}(s)ds\nonumber\\
&=1-Q\bigg(\frac{\epsilon-\mu_{\Delta1}}{\sigma_{\Delta1}}\bigg)+Q\bigg(\frac{\epsilon-\mu_{\Delta0}}{\sigma_{\Delta0}}\bigg).
\end{align}
By differentiating $\varepsilon(\epsilon)$ with respect to $\epsilon$ and equating the result to 0, we obtain
\begin{align}
\label{eq:equal}
\exp\bigg(-\frac{(\epsilon-\mu_{\Delta0})^2}{2\sigma^2_{\Delta0}}\bigg)=\frac{\sigma_{\Delta0}}{\sigma_{\Delta1}}\exp\bigg(-\frac{(\epsilon-\mu_{\Delta1})^2}{2\sigma^2_{\Delta1}}\bigg).
\end{align}
From \eqref{eq:deltav0}, \eqref{eq:deltav1}, we can find that $\sigma^2_{\Delta0}>\sigma^2_{\Delta1}$ holds, and then, we derive that
\begin{align}
\sqrt{\sigma^2_{\Delta0}\sigma^2_{\Delta1}\mu^2_{\Delta1}+2\sigma^2_{\Delta0}\sigma^2_{\Delta1}(\sigma^2_{\Delta0}-\sigma^2_{\Delta1})\ln\frac{\sigma_{\Delta0}}{\sigma_{\Delta1}}}>0.
\end{align}
After some algebraic manipulations, we obtain the optimal detection threshold $\epsilon^\star$ in \eqref{eq:threshold} and complete the proof.
\ifCLASSOPTIONcaptionsoff
  \newpage
\fi



\bibliographystyle{IEEEtran}
\bibliography{IEEEabrv,sigproc} 
\end{document}